\documentclass{elsart}
\usepackage{graphicx, amssymb}
\journal{Phys. Lett. A}

\begin{document}

\begin{frontmatter}



\title{Exact solution of the spin-1/2 Ising model on the Shastry-Sutherland 
       (orthogonal-dimer) lattice\thanksref{VEGA}}
\thanks[VEGA]{This work was financially supported under the grant VEGA 1/2009/05.}

\author{Jozef Stre\v{c}ka}
\ead{jozkos@pobox.sk}
\address{Department of Theoretical Physics and Astrophysics, 
Faculty of Science, \\ P. J. \v{S}af\'{a}rik University, Park Angelinum 9,
040 01 Ko\v{s}ice, Slovak Republic}

\begin{abstract}
A star-triangle mapping transformation is used to establish an exact correspondence between the spin-1/2 Ising model on the Shastry-Sutherland (orthogonal-dimer) lattice and respectively, the spin-1/2 Ising model on 
a bathroom tile (4-8) lattice. Exact results for the critical temperature and spontaneous magnetization are obtained and compared with corresponding results on the regular Ising lattices.
\end{abstract}


\end{frontmatter}


\section{Introduction}

Over the last six decades, much effort has been devoted to determine a criticality and other statistical properties of various lattice-statistical models, which would enable a deeper understanding of order-disorder phenomena in solids. The planar Ising model takes a prominent place in the equilibrium statistical mechanics
from this point of view as it represents a rare example of exactly solvable lattice-statistical model since Onsager's pioneering work \cite{onsager}. It is noteworthy that early exact studies of the planar Ising model have been performed on regular lattices - square \cite{onsager}, triangular and honeycomb \cite{triang}, having all sites as well as all bonds equivalent. Exact calculations performed on these lattices revealed 
the universal critical behavior supporting the universality hypothesis, in fact, all three regular lattices exhibit a second-order phase transition characterized by the same set of critical indices. 

A remarkable generalization has been achieved when the planar Ising models with non-crossing bonds were 
shown to be, at least in principle, exactly solvable within the Pfaffian method \cite{green}. In spite of 
the universality in their critical behavior, it is still of great research interest to calculate accurately the critical point and other statistical properties for a specific lattice with particular interactions. 
Exact studies of the Ising model on the Union Jack (centered square) lattice proved, for instance, a possible existence of reentrant transitions in a presence of competing interactions \cite{union}. It should be mentioned, however, that a precise treatment of planar Ising models is usually connected with considerable difficulties related to sophisticated mathematical methods used. On the other hand, adopting the well-known formalism based on mapping transformation technique \cite{MT} largely simplifies this treatment. By 
the use of mapping, exact solutions of the Ising model on a pentagonal lattice \cite{urumov} and two 
topologically different 4-6 lattices \cite{46a}-\cite{46b} have recently been derived. 

From a topological viewpoint, the Archimedean (uniform) lattices attract an appreciable attention because 
they represent infinite tessellation in which all sites are equivalent \cite{GS}. In addition to a subset of 
three regular lattices, the other eight Archimedean lattices being irregular since their tilings consist 
of two or more regular polygons. Although the Ising model on some of these 'semi-regular' lattices
have exactly been treated long ago, e.g. on Kagom\'e \cite{kagome}, expanded Kagom\'e (3-12) \cite{312} 
or bathroom tile (4-8) \cite{48} lattices, some of them seem not to have been dealt in the literature so far. 
The primary goal of this work is to find out an accurate solution for the spin-1/2 Ising model on the Shastry-Sutherland lattice \cite{SS} (to be further abbreviated as SSL) schematically shown on right-hand
side of Fig. 1. The SSL is five-coordinated Archimedean lattice formed by square plaquettes each 
surrounded by four triangular plaquettes. Consequently, there are two non-equivalent bonds in the SSL 
which may be a common edge either of two triangles, or a triangle and a square. Notice that the lattice topology of SSL can also be viewed as a planar network constituted by mutually orthogonal neighboring 
dimers (see Fig. 1). It is worthwhile to mention that a theoretical investigation of the Ising model 
on the SSL is interesting from the academic point of view as well as in connection with a recent 
experimental observation of insulating molecular-based compounds SrCu$_2$(BO$_3$)$_2$ \cite{SSL-Cu}, Nd$_2$BaZnO$_5$ \cite{SSL-Nd} and [Ni(mettn)(N$_3$)] \cite{SSL-Ni} whose magnetic structure matches 
a requirement of the SSL topology. 

The organization of this paper is as follows. In the next section we derive an exact solution 
for the spin-1/2 Ising model on the SSL within an exact mapping to a corresponding bathroom tile (4-8) lattice. Exact results for the critical temperature and spontaneous magnetization are studied particularly
in Section 3. Finally, some concluding remarks are given in Section 4.

\section{Model and its exact solution}

Let us begin by considering the anisotropic bathroom tile (4-8) lattice schematically illustrated on the left-hand side of Fig. 1. While all sites of the 4-8 lattice are equivalent, having coordination number three, the lattice is irregular due to the non-equivalency of its bonds. In this respect, two different coupling parameters are assigned to the bonds within square plaquettes ($R_2$) and respectively, to the links between them ($R_1$). Another important topological feature of the 4-8 lattice is that it belongs to 'loose-packed' lattices,  i.e. it can be divided into two equal sub-lattices $A$ and $B$ in such a way that all nearest neighbors of $A$-sites belong to the sub-lattice $B$ and vice versa. To emphasize this topological feature, the $A$-sites ($B$-sites) are displayed in Fig. 1 as empty (solid) circles. On assumption that all sites are occupied by spin-1/2 atoms, the Ising Hamiltonian defined upon the underlying 4-8 lattice reads:
\begin{eqnarray}
{\mathcal H}_{4-8} = - R_1 \sum_{(i,j) \subset \mathcal L}^{N} S_{i}^{A} S_{j}^{B} 
                 - R_2 \sum_{(i,j) \subset \mathcal S}^{2N} S_{i}^{A} S_{j}^{B},     
\label{H48}
\end{eqnarray}
where $S^{\alpha}_i = \pm 1$ ($\alpha = A, B$) is the Ising spin located at $i$th lattice point, its 
superscript specifies the sub-lattice to which it belongs, $2N$ denotes the total number of lattice 
sites and the summations run over two sets of bonds, the bonds within the square plaquettes 
and the links between them, respectively. 

It is of principal importance that the 4-8 lattice can be divided into two interpenetrating sub-lattices. 
Owing to this fact, a summation over the spins of one sub-lattice (say sub-lattice $B$) can be performed 
before summing over the spins of another sub-lattice (sub-lattice $A$). Accordingly, the partition function 
of 4-8 lattice can be rewritten in this useful form:
\begin{eqnarray}
{\mathcal Z}_{4-8} = \sum_{\{S^{A} \}} \prod_{i=1}^{N} \sum_{S^{B}_{i} = -1}^{+1}
                                       \exp[S^B_i (L_1 S^A_j + L_2(S^A_k + S^A_l)],     
\label{Z48}
\end{eqnarray}
where $L_1 = R_1/(k_{\mathrm B} T)$ and $L_2 = R_2/(k_{\mathrm B} T)$ are effective coupling parameters, 
$k_{\mathrm B}$ is Boltzmann's constant and $T$ the absolute temperature. Further, the product is over all $B$-sites and the symbol $\sum_{\{S^{A} \}}$ means a summation over all possible spin configurations on the sub-lattice $A$. Apparently, this form justifies performance of the familiar star-triangle mapping transformation \cite{MT}:
\begin{eqnarray}
\sum_{S^{B}_{i} = -1}^{+1} \exp[S^B_i (L_1 S^A_j &+& L_2(S^A_k + S^A_l)] \nonumber \\
        = A \exp[\frac{K_1}{2} S^A_k S^A_l &+& K_2 S_j^A(S^A_k + S^A_l)].     
\label{ST}
\end{eqnarray}
A physical meaning of the mapping (\ref{ST}) is to remove all interaction parameters associated with 
the $B$-site located at $i$th lattice point and to replace them by new effective coupling parameters 
$K_1 = J_{\rm intra}/(k_{\mathrm B} T)$ and $K_2 = J_{\rm inter}/(k_{\mathrm B} T)$ between the remaining $A$-sites. When the mapping (\ref{ST}) is applied to all $B$-sites, the 4-8 lattice is then mapped on the corresponding SSL (Fig. 1). In what follows, we will refer to new effective interactions $K_1$ and $K_2$ 
of the SSL as to the intra- and inter-dimer couplings, respectively. Note that the mapping parameters $A$, $K_1$ and $K_2$ are unambiguously given by the transformation (\ref{ST}), which must be valid for any configuration of $S_j^A$, $S^A_k$, $S^A_l$ spins. Hence, 
\begin{eqnarray}
A &=& 2 [\cosh(L_1 + 2L_2) \cosh(L_1 - 2L_2) \cosh^2(L_1)]^{1/4},  
\label{MP1} \\
K_1 &=& \frac12 \ln \Bigl[ \frac{\cosh(L_1 + 2L_2) \cosh(L_1 - 2L_2)}{\cosh^2(L_1)} \Bigr], 
\label{MP2} \\
K_2 &=& \frac14 \ln \Bigl[ \frac{\cosh(L_1 + 2L_2)}{\cosh(L_1 - 2L_2)} \Bigr]. 
\label{MP3}
\end{eqnarray}
It is worthy to notice that a substitution of the transformation Eq. (\ref{ST}) into the formula 
(\ref{Z48}) establishes a relationship between the partition function of the 4-8 lattice and 
its corresponding partition function of the SSL lattice:
\begin{eqnarray}
{\mathcal Z}_{4-8} (L_1, L_2) = A^N {\mathcal Z}_{SSL} (K_1, K_2).
\label{PF}
\end{eqnarray}
Of course, similar mapping relations can also be established for other important quantities such as 
Gibbs free energy, internal energy, magnetization, correlation functions, specific heat, etc.
For instance, with the aid of Eq. (\ref{PF}) it can readily be proved that a magnetization of the 
4-8 lattice, characterized by the effective interactions $L_1$ and $L_2$, directly equals to a 
magnetization of the SSL with the coupling parameters $K_1$ and $K_2$ given by (\ref{MP2})-(\ref{MP3}):   
\begin{eqnarray}
m_{4-8} (L_1, L_2) =  m_{SSL} (K_1, K_2).
\label{mag}
\end{eqnarray}
 
Now, we will find the {\it inverse transformation} in order to express an exact solution of 
the SSL in terms of the known exact solution of the 4-8 lattice \cite{48}. After elementary 
calculations one finds inverse transformation formulae:
\begin{eqnarray}
L_1 &=& \frac12 \ln \biggl[ \frac{\sqrt{({\mathrm e}^{K_1}\cosh 2K_2)^2-1} + {\mathrm e}^{K_1}\sinh 2K_2}
                           {\sqrt{({\mathrm e}^{K_1}\cosh 2K_2)^2-1} - {\mathrm e}^{K_1}\sinh 2K_2} \biggr], 
\label{IP1} \\                               
L_2 &=& \frac12 \ln \biggl[ {\mathrm e}^{K_1} \cosh 2K_2 \pm \sqrt{({\mathrm e}^{K_1}\cosh 2K_2)^2-1} \biggr].
\label{IP2}
\end{eqnarray}
It should be stressed, however, that a validity of the mapping relation (\ref{IP1}) necessitates $K_1 \geq 0$, i.e. the mapping between the SSL and 4-8 lattice holds for the ferromagnetic intra-dimer interaction only. On the other hand, a sign ambiguity in the mapping relation (\ref{IP2}) is of less importance, since basic features of the 4-8 Ising lattice such as critical frontiers, partition function or Gibbs free energy remain invariant under the sign change transforming merely $L_2 \to - L_2$. With regard to this, the 
partition function and magnetization of spin-1/2 Ising model on the SSL lattice can immediately be 
calculated from the corresponding quantities of the 4-8 lattice \cite{48} with the help of Eqs. (\ref{PF}) 
and (\ref{mag}) when exploiting the transformation formulae (\ref{IP1})-(\ref{IP2}). 

\section{Results and Discussion}

Let us proceed to a discussion of the most interesting numerical results. For simplicity, we shall confine 
our attention to a particular case when both intra- as well as inter-dimer coupling constants of 
the SSL are assumed to be ferromagnetic. It should be mentioned, however, that the mapping relations (\ref{IP1})-(\ref{IP2}) are responsible merely for a sign change of the coupling parameter 
$L_1$ when transforming $K_2 \to -K_2$, what means that the presented results for critical boundaries show 
the critical temperature of SSL with a ferromagnetic intra-dimer and an antiferromagnetic inter-dimer interaction as well.

The critical temperatures of the anisotropic square, SSL and triangular Ising lattices are displayed 
in Fig. 2 as a function of the ratio between relevant interaction parameters. The interaction parameters $J_{\rm intra}$ and $J_{\rm inter}$ are defined for all three Ising lattices in Fig. 3 representing them 
as solid and dashed lines, respectively. For better orientation, Fig. 2A (Fig. 2B) illustrates critical boundaries of the anisotropic Ising lattices when $J_{\rm intra}$ is stronger (weaker) than $J_{\rm inter}$. As one can see from Fig. 2A, the critical temperatures gradually increase as $J_{\rm inter}/J_{\rm intra}$ strengthens as many as their maximum values corresponding to isotropic lattices are reached. Apart from this 
rather trivial finding it is quite remarkable to ascertain that the critical temperature of five-coordinated SSL is lower than the one of four-coordinated square lattice when $J_{\rm inter}/J_{\rm intra} < 0.458$.
This difference is with a high certainty attributable to different relative occurrencies of the 
$J_{\rm intra}$- and $J_{\rm inter}$-bonds, which are in the ratio 1:1 for the square lattice, whereas 
they are merely in the ratio 1:4 for the SSL. Another interesting fact to observe here is a significant departure between the critical temperatures of both regular Ising lattices and respectively, the SSL in a weak-$J_{\rm inter}$ region, where the critical temperature of SSL does not seem to follow a steep 
logarithmic increase observed for regular lattices. 

As far as a strong-$J_{\rm inter}$ limit is considered (Fig. 2B), an increasing strength of the ratio 
$J_{\rm intra}/J_{\rm inter}$ is then responsible for almost a linear increase of the critical temperature 
of triangular and SSL lattices. Dependences drawn for critical temperatures of these lattices starting 
both from the critical point of isotropic square lattice and achieve their maximum values when approaching 
the isotropic limit. Nevertheless, the best linear fits of both the curves reveal that a linear term determining a slope of critical curve is for the triangular lattice roughly twice as large as for the SSL. This could be understood as a consequence of two times greater number of $J_{\rm intra}$-bonds added to a simple square lattice when constructing the anisotropic triangular lattice (see Fig. 3). Finally, 
it appears worthwhile to remark that our numerical value for the critical point of isotropic SSL 
($k_{\rm B} T_c / J = 2.926261$) correctly reproduces a critical value reported by Thompson and Wardrop  \cite{tcSSL} several years ago by the use of Vdovichenko's combinatorial approach. 

To complete our analysis, the temperature dependence of spontaneous magnetization is depicted in Fig. 4 
for several values of the ratio $J_{\rm inter}/J_{\rm intra}$. As could be expected, the spontaneous magnetization varies very abruptly slightly below its critical temperature as $\sim A \varepsilon^{1/8}$, where $A$ is termed the critical amplitude and $\varepsilon \equiv (T_c-T)/T_c$ being the fractional deviation of the temperature from its critical value. This magnetization-type singularity, which is often called also 
as an algebraic branch point, can serve as a convincing evidence that the anisotropic SSL falls into a standard Ising universality class independently of the $J_{\rm inter}/J_{\rm intra}$ strength. 

\section{Conclusion}

Exact results for the critical temperature and spontaneous magnetization of the spin-1/2 Ising model 
on the anisotropic SSL are obtained by means of the star-triangle transformation. The star-triangle 
mapping largely simplifies an exact treatment of the anisotropic SSL in that it relates these exact results 
to the ones of corresponding 4-8 lattice. Although the investigation of spin-1/2 Ising model on the SSL 
has primarily been stimulated by a recent exploration of molecular-based compounds having the magnetic structure of SSL topology \cite{SSL-Cu}-\cite{SSL-Ni}, its significance should also be viewed in connection with manifold applications of the Ising model in seemingly diverse research areas. Actually, there are 
various physical motivations to examine the lattice-statistical Ising models which proved their usefulness 
in the realm of statistical physics as suitable models for investigating a phase separation in liquids, lattice gas models capturing liquid-gas transitions, lattice models of alloys, models explaining 
cooperative phenomena in various biological and chemical systems, and so on.   



\newpage
\begin{large}
\textbf{Figure captions}
\end{large}

\begin{itemize}

\item [Fig. 1]
A schematic representation of the mapping scheme. The spatially anisotropic bathroom tile (4-8) lattice, formally divided into two equal sub-lattices $A$ (empty circles) and $B$ (solid circles), is mapped onto 
the anisotropic Shastry-Sutherland (orthogonal-dimer) lattice with the help of star-triangle transformation 
removing all $B$-type sites. 

\item [Fig. 2]
Exact critical temperatures of the anisotropic square, SSL and triangular Ising lattices in dependence 
on the ratio $J_{\rm inter}/J_{\rm intra}$ (Fig. 2A) and $J_{\rm intra}/J_{\rm inter}$ (Fig. 2B), respectively.

\item [Fig. 3]
A schematic representation of the anisotropic square (Fig. 3A), SSL (Fig. 3B) and triangular (Fig. 3C) lattices. The coupling $J_{\rm intra}$ ($J_{\rm inter}$) is represented by solid (dashed) line style.

\item [Fig. 4]
The temperature variation of magnetization normalized per one lattice site of the SSL when the ratio 
$J_{\rm inter}/J_{\rm intra}$ changes. Temperature is scaled in the dimensionless units.

\end{itemize} 


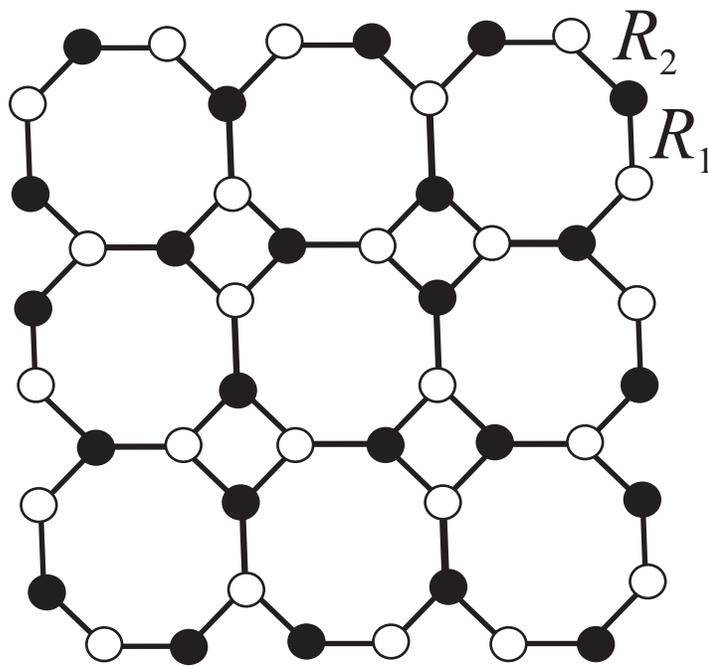
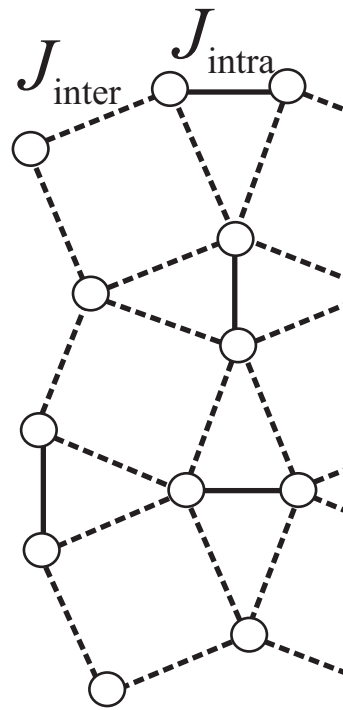
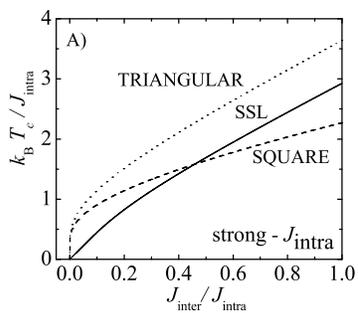
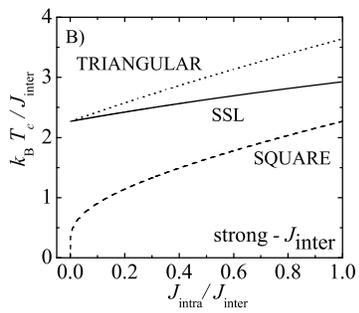
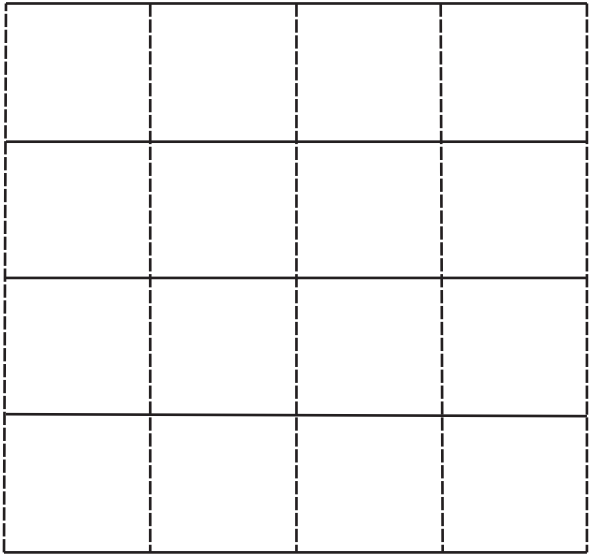
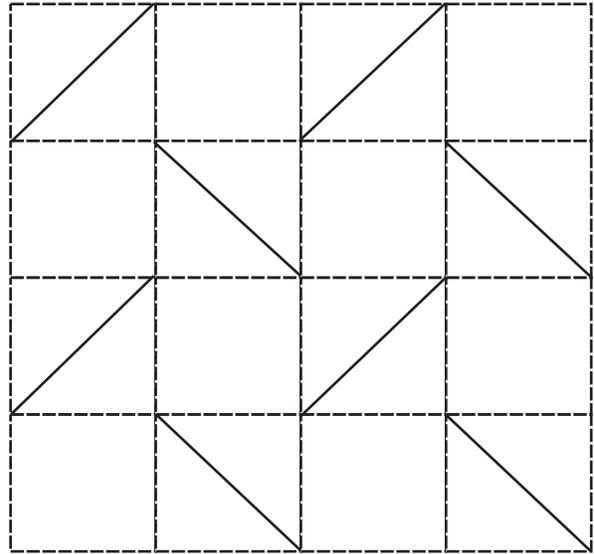
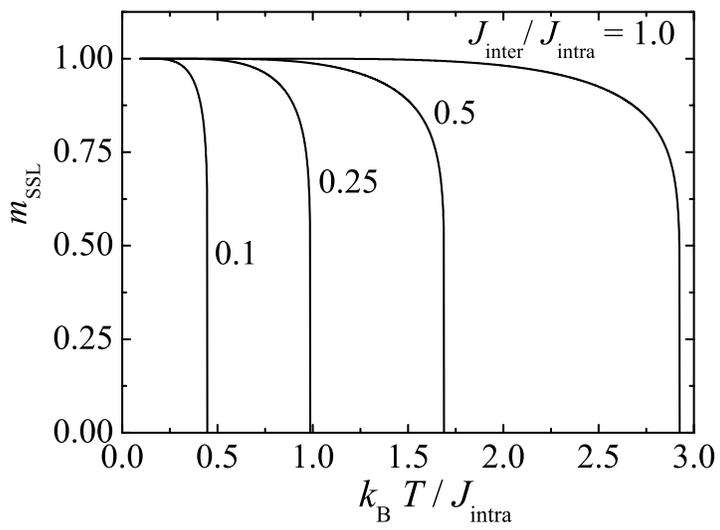
\end{document}